\documentclass[fleqn,twoside]{article}
\usepackage{espcrc2}
\usepackage{epsfig}
\usepackage{latexsym}
\usepackage{amssymb}
\usepackage{amsfonts}

\readRCS
$Id: espcrc2.tex,v 1.2 2004/02/24 11:22:11 spepping Exp $
\usepackage{graphicx}
\usepackage[figuresright]{rotating}

\newcommand{\AmS}{{\protect\the\textfont2
  A\kern-.1667em\lower.5ex\hbox{M}\kern-.125emS}}

\title{Dimensional recurrence relations: \\ an easy way to evaluate
higher orders of expansion in $\ep$}

\author{Roman N.~Lee\address{Budker Institute of Nuclear Physics and
 Novosibirsk State University,  Novosibirsk 630090, Russia},
Alexander V.~Smirnov\address{Scientific Research Computing Center of
Moscow State University, Moscow 119992, Russia} and
Vladimir A.~Smirnov\address{Nuclear Physics Institute of
  Moscow State University,
  Moscow 119992, Russia}\thanks{Talk given at the International
  Workshop `Loops and Legs in
  Quantum Field Theory' (April 25--30, 2010, W\"orlitz, Germany).}
}

\newcommand{\ep}{\varepsilon}
\def\e{{\rm e}}
\newcommand{\be}{\begin{equation}}
\newcommand{\ee}{\end{equation}}
\newcommand{\bea}{\begin{eqnarray}}
\newcommand{\eea}{\end{eqnarray}}

\newcommand{\nn}{\nonumber}

\begin{document}

\begin{abstract}
Applications of a method recently suggested by one of the authors (R.L.)
are presented. This method is based on the use of dimensional recurrence
relations and analytic properties
of Feynman integrals as functions of the parameter of dimensional regularization,
$d$.
The method was used to obtain analytical expressions for two
missing constants in the $\ep$-expansion of the most complicated
master integrals contributing to the three-loop massless quark and gluon form factors
and thereby present the form factors in a completely analytic form.
To illustrate its power we present, at transcendentality weight seven, the next order of
the $\ep$-expansion of one of the corresponding most complicated master integrals.
As a further application, we present three previously unknown terms of the expansion
in $\ep$ of the three-loop non-planar massless propagator diagram.
Only multiple $\zeta$ values at integer points are present in our result.
\vspace{1pc}
\end{abstract}

\maketitle

\sloppy

\section{Introduction}

Quite recently a new method \cite{Lee:2009dh} of evaluating Feynman integrals was suggested.
It is based on the use of dimensional recurrence relations \cite{Tarasov1996}
and analytic properties
of Feynman integrals as functions of the parameter of dimensional regularization,
$d$. Here we are going to describe new results of its application.
In the next section we give a summary of the evaluation of the two previously
analytically unknown coefficients of transcendentality weight six in the
$\ep$-expansion of two master integrals contributing to the three-loop massless
form factors \cite{Lee:2010cg}.
We also present, at transcendentality weight seven, the next order of
the $\ep$-expansion of one of the corresponding most complicated master integrals.
In Section~3,  we report on the evaluation of higher orders of expansion in $\ep$
of the three-loop non-planar massless propagator diagram and, in Section~4, we discuss
our results.

\section{Evaluating master integrals for \\
three-loop form factors}

The integration-by-part reduction reduces the problem to the calculation
of a small number of master integrals. All the master integrals apart from three most
complicated master integrals contributing to the three-loop massless form factors have been
analytically evaluated in \cite{3lff1,3lff2}.
About one year ago, one of the three
most complicated master integrals (called $A_{9,1}$ in \cite{3lff1,3lff2,3lff3,3lff4})
and the pole
parts of $A_{9,4}$ and $A_{9,2}$  shown in Figs.~1 and~2
 \begin{figure}[htb]
  \begin{center}
    \includegraphics[width=.4\textwidth]{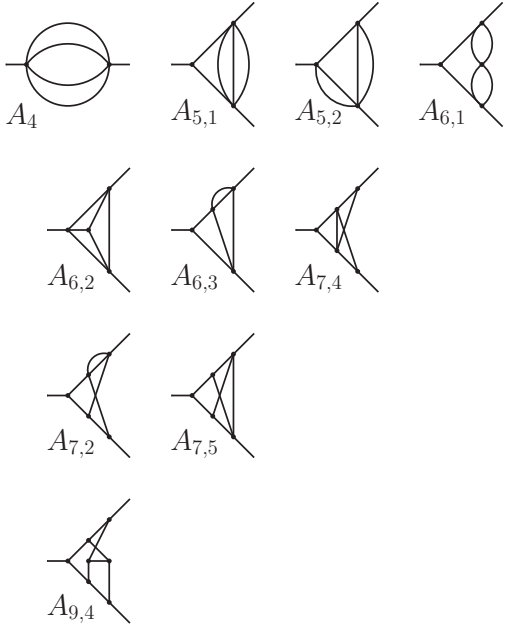}
   \caption[]{Master integrals for $A_{9,4}$.}
  \end{center}
\end{figure}
\begin{figure}[htb]
  \begin{center}
    \includegraphics[width=.4\textwidth]{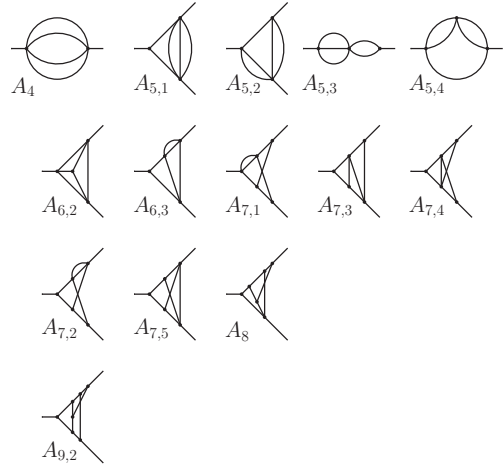}
   \caption[]{Master integrals for $A_{9,2}$.}
  \end{center}
\end{figure}
were evaluated analytically, while the $\ep^0$ parts of
$A_{9,4}$ and $A_{9,2}$ were evaluated numerically --- see \cite{3lff3,3lff4}.

In \cite{Lee:2010cg}, the two missing ingredients, i.e. the finite parts of
$A_{9,4}$ and $A_{9,2}$, were evaluated analytically.
According to the method of \cite{Lee:2009dh}
it is necessary, before evaluating $A_{9,4}$ and $A_{9,2}$,
to know
all lower master integrals. They are shown in the same figures.
Four rows of diagrams in each figure correspond to complexity levels 0, 1, 2 and 3.
Details of the calculation can be found in \cite{Lee:2009dh}.
Here are the corresponding results:
\begin{eqnarray}
A_{9,4}(4-2\ep) =\e^{-3\gamma_E \ep }\biggl\{ -\frac{1}{9\ep^6}-\frac{8}{9\ep^5}
&& \nn \\ &&  \hspace*{-62mm}
+\left[ 1+\frac{43 \pi ^2}{108}\right]\frac{1}{\ep^4}
+\left[\frac{109 \zeta (3)}{9}+\frac{14}{9}
+\frac{53 \pi ^2}{27} \right]\frac{1}{\ep^3}
\nn \\ &&  \hspace*{-62mm}
+\left[ \frac{608 \zeta (3)}{9}-17-\frac{311 \pi ^2}{108}
-\frac{481 \pi^4}{12960}\right]\frac{1}{\ep^2}
\nn \\ &&  \hspace*{-62mm}
+\left[ -\frac{949 \zeta (3)}{9}-\frac{2975 \pi ^2 \zeta (3)}{108}
+\frac{3463\zeta (5)}{45} \right.
\nn \\ &&  \hspace*{-62mm}
\left.
+84+\frac{11 \pi ^2}{18}
+\frac{85 \pi ^4}{108}\right]\frac{1}{\ep}
\nn \\ &&  \hspace*{-62mm}
+\left[\frac{434 \zeta (3)}{9}-\frac{299 \pi ^2 \zeta (3)}{3}
-\frac{3115 \zeta(3)^2}{6}+\frac{7868 \zeta (5)}{15}
\right. \nn \\ &&  \hspace*{-62mm}
\left. -339+\frac{77 \pi ^2}{4}
-\frac{2539 \pi^4}{2592}-\frac{247613 \pi ^6}{466560} \right] +O(\ep)\biggr\}
\;;
\nn
\label{A94}
\end{eqnarray}
\begin{eqnarray}
A_{9,2}(4-2\ep) = \e^{-3\gamma_E \ep }\biggl\{-\frac{2}{9\ep^6}-\frac{5}{6\ep^5}
&& \nn \\ &&  \hspace*{-62mm}
+\left[ \frac{20}{9}+\frac{17 \pi ^2}{54}\right]\frac{1}{\ep^4}
\nn \\ &&  \hspace*{-62mm}
+\left[\frac{31 \zeta (3)}{3}-\frac{50}{9}
+\frac{181 \pi ^2}{216}  \right]\frac{1}{\ep^3}
\nn \\ &&  \hspace*{-62mm}
+\left[\frac{347 \zeta (3)}{18}+\frac{110}{9}-\frac{17 \pi ^2}{9}
+\frac{119 \pi ^4}{432} \right]\frac{1}{\ep^2}
\nn \\ &&  \hspace*{-62mm}
 +\left[-\frac{514 \zeta (3)}{9}-\frac{341 \pi ^2 \zeta (3)}{36}
+\frac{2507 \zeta(5)}{15}\right.
\nn \\ &&  \hspace*{-62mm}
\left.
-\frac{170}{9}+\frac{19 \pi ^2}{6}
+\frac{163 \pi ^4}{960} \right]\frac{1}{\ep}
\nn \\ &&  \hspace*{-62mm}
+\left[\frac{1516 \zeta (3)}{9}-\frac{737 \pi ^2 \zeta (3)}{24}-29 \zeta (3)^2
+\frac{2783\zeta (5)}{6}
\right. \nn \\ &&  \hspace*{-62mm}
\left.-\frac{130}{9}+\frac{\pi ^2}{2}
-\frac{943 \pi ^4}{1080}+\frac{195551 \pi ^6}{544320}\right] +O(\ep)\biggr\}
\;.
\label{A92}
\nn
\end{eqnarray}

An independent calculation of the form factors was performed quite recently
\cite{Gehrmann:2010ue} and the agreement with the previous results was established.
Motivated by a future four-loop calculation the authors calculated also the
subleading $O(\ep)$ terms for the fermion-loop type  contributions.
We think that the analytic calculation of the whole $O(\ep)$ part of
the form factors is feasible at the moment. To illustrate this possibility
we have calculated the $O(\ep^2)$ order of
the $\ep$-expansion of one of the corresponding most complicated master integrals,
$A_{9,1}$, which is of transcendentality weight seven:
\begin{eqnarray}
A_{9,1}(4-2\ep) = e^{-3\gamma_E \ep }\biggl\{
\frac{1}{18 \ep^5}-\frac{1}{2\ep^4}
&& \nn \\ &&  \hspace*{-60mm}
+\frac{1}{\ep^3}\left(\frac{53}{18}+\frac{29 \pi^2}{216}\right)
\nn \\ &&  \hspace*{-60mm}
+\frac{1}{\ep^2}
\left(\frac{ 35 \zeta(3)}{18}-\frac{29}{2}-\frac{149 \pi ^2}{216}\right)
\nn \\ &&  \hspace*{-60mm}
+\frac{1}{\ep}
\left(-\frac{307 \zeta(3)}{18}+\frac{129}{2}+\frac{139 \pi ^2}{72}
+\frac{5473 \pi^4}{25920}\right)
\nn \\ &&  \hspace*{-60mm}
 +\left(\frac{793 \zeta (5)}{10}+\frac{871 \pi^2\zeta (3)}{216}+\frac{1153 \zeta (3)}{18}
 \right.
\nn \\ && \hspace*{-60mm}
 \left.
 -\frac{3125 \pi^4}{5184}-\frac{19 \pi ^2}{8}-\frac{537}{2}\right)
\nn \\ &&  \hspace*{-60mm}
+\ep
   \left(-\frac{287 \zeta (3)}{2}+\frac{2969 \pi ^2 \zeta(3)}{216}
+\frac{5521 \zeta (3)^2}{36}
\right.
\nn \\ &&  \hspace*{-60mm}
\left.
 -\frac{8251 \zeta(5)}{30}
   +\frac{2133}{2}
   -\frac{97 \pi ^2}{8}
\right.
\nn \\ &&  \hspace*{-60mm}
\left.
+\frac{4717 \pi^4}{28115}+\frac{761151 \pi ^6}{186624}\right)
\nn \\ &&  \hspace*{-60mm}
+\ep^2
\left(\frac{195 \zeta (3)}{2}-\frac{5887 \pi ^2 \zeta (3)}{72}
+\frac{138403\pi ^4 \zeta (3)}{25920}
\right.
\nn \\ &&  \hspace*{-60mm}
\left.
+\frac{799 \zeta (3)^2}{4}
+\frac{22487 \zeta(5)}{30}
-\frac{11987 \pi ^2 \zeta (5)}{10115}
\right.
\nn \\ &&  \hspace*{-60mm}
\left.
+\frac{228799 \zeta(7)}{126}-\frac{8181}{2}+\frac{969 \pi ^2}{8}
\right.
\nn \\ &&  \hspace*{-60mm}
\left.
-\frac{1333 \pi^4}{320}-\frac{4286603 \pi ^6}{6531840}\right)
+\ldots \biggr\} \;.
\label{A91ep2} \nn
\end{eqnarray}

\section{Non-planar massless propagator diagram}

It is also natural to try to apply the same method to evaluate
the three previously
analytically unknown coefficients of transcendentality weight six in the
$\ep$-expansion of three master integrals contributing to the three-loop
static quark potential
\cite{Smirnov:2007zz,Smirnov:2008pn,Smirnov:2008ay,Smirnov:2009fh,Anzai:2009tm,Smirnov:2010zc}.
These are $I_{11},I_{16},I_{18}$ in Fig.~3.
\begin{figure}[h]
  \begin{center}
    \includegraphics[width=.5\textwidth]{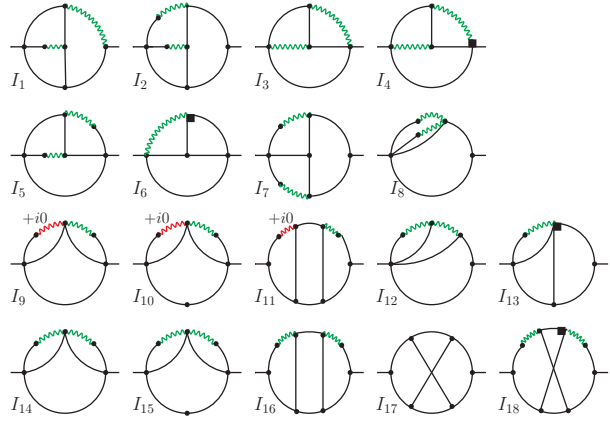}
   \caption[]{Most complicated master integrals for the three-loop static quark potential.}
  \end{center}
\end{figure}

This work is in progress \cite{LSS1}. Here we would like to present, as a by-product
of this activity,
new results for one of the master integrals which are lower than $I_{18}$.
Let us consider $I_{17}$ which is a master integral for three-loop massless propagator
integrals. The famous value $20\zeta(5)$ at $d=4$ is known for a long time
\cite{CKT}. The linear term in the $\ep$-expansion was obtained in \cite{Kaz}.
The quadratic term was recently evaluated \cite{Bec}.
To illustrate the power of the present method we have evaluated the three next terms
so that we have a result up to $\ep^5$.

\begin{figure}[htb]
  \begin{center}
    \includegraphics[width=.4\textwidth]{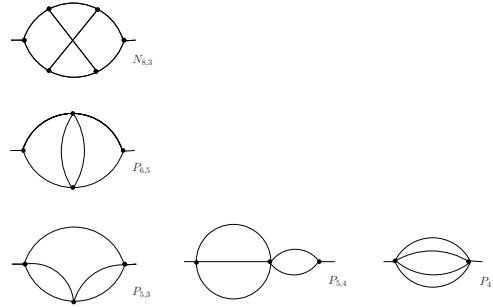}
\caption[]{Master integrals for $I_{17}$.}
\label{n83}
\end{center}
\end{figure}

The diagram itself and the corresponding four lower diagrams are shown in Fig.~\ref{n83}
The labelling for $I_{17}\equiv N_{8,3}$ and its lower master integrals is taken from
the future publication \cite{LSS1}.
This master integral itself has complexity level 2,
the master integral $P_{6,5}$ has complexity level 1, while the other three master integrals
can be expressed in terms of gamma functions at general $d$.

The lowering recurrence relation for this integral provides the following expression
of $N_{8,3}^{(d+2)}$ in terms of master integrals in $d$ dimensions:
\bea
N_{8,3}^{(d+2)}=
\frac{(d-4) }{8 (d-2) (d-1) (2 d-7) (2 d-5)}N_{8,3}^{(d)}
&& \nn \\ &&  \hspace*{-77mm}
+\frac{4 \left(5 d^2-28 d+38\right)}
{(d-4)^2 (d-2) (d-1) (2 d-5)} P_{5,3}^{(d)}
\nn \\ &&  \hspace*{-77mm}
+\left[
4 (d-4) (d-2) (d-1) (2 d-7) (2 d-5) 
\right]^{-1}
\nn \\ &&  \hspace*{-77mm}
\times(3 d-8)^{-1}\left(37 d^3-313 d^2+858 d-752\right) P_{6,5}^{(d)}
\nn \\ &&  \hspace*{-77mm}
-\left(43 d^4-478 d^3+1963 d^2-3530 d+2352\right)
\nn \\ &&  \hspace*{-77mm}
\times\left[
2 (d-4)^2 (d-3) (d-2) (d-1) 
\right]^{-1}
\nn \\ &&  \hspace*{-77mm}
\times 
\left[(2 d-7) (2 d-5)\right]^{-1}
P_{5,4}^{(d)}
\nn \\ &&  \hspace*{-77mm}
-\left[
(d-4)^3 (d-3)^2 (d-2) 
\right]^{-1}
\nn \\ &&  \hspace*{-77mm}
\times
\left[(d-1) (2 d-7) (3 d-8)\right]^{-1}
\nn \\ &&  \hspace*{-77mm}
\times\left(401 d^6-7251 d^5+54491 d^4-217784 d^3
\right.
\nn \\ &&  \hspace*{-77mm}
\left.
+486264 d^2-581248 d+287232\right) P_4^{(d)}
\;.\nn
\eea

We obtain the following result to order $\ep^5$:
\bea
I_{17}=\frac{\e^{-3\gamma_E \ep}}{1-2\ep}\biggl\{
20 \zeta (5)
+\ep \left(68 \zeta (3)^2+\frac{10 \pi^6}{189}\right)
&& \nn \\ &&  \hspace*{-74mm}
 +\ep^2 \left(\frac{34 \pi ^4 \zeta (3)}{15}-5 \pi ^2 \zeta(5)+450 \zeta (7)\right)
\nn \\ &&  \hspace*{-74mm}
+\ep^3 \left(-\frac{9072}{5}\zeta(5,3)-2588 \zeta (3) \zeta (5)
\right.
\nn \\ &&  \hspace*{-74mm}
\left.
-17 \pi ^2 \zeta (3)^2
+\frac{6487\pi ^8}{10500}\right)
\nn \\ &&  \hspace*{-74mm}
+\ep^4 \left(-\frac{4897 \pi ^6 \zeta(3)}{630}-\frac{6068 \zeta (3)^3}{3}
+\frac{13063 \pi^4 \zeta(5)}{120}
\right.
\nn \\ &&  \hspace*{-74mm}
\left.
-\frac{225 \pi ^2 \zeta (7)}{2}
+\frac{88036 \zeta(9)}{9}\right)
\nn \\ &&  \hspace*{-74mm}
+\ep^5 \left(\frac{2268}{5} \pi ^2 \zeta(5,3)+42513\zeta(8,2)
\right.
\nn \\ &&  \hspace*{-74mm}
\left.
-145328 \zeta (3) \zeta (7)-73394 \zeta (5)^2
+647 \pi ^2 \zeta (3) \zeta (5)
\right.
\nn \\ &&  \hspace*{-74mm}
\left.
-\frac{11813 \pi ^4 \zeta (3)^2}{120}+\frac{28138577 \pi^{10}}{9355500}\right)
+\ldots
\biggr\}\;,
\nn
\eea
where $\zeta(n,m)$ are multiple zeta values (see, e.g., \cite{BBV}).
Observe that the pulling out the standard prefactor $1/(1-2\ep)$ in results
for massless propagator integrals (see, e.g., \cite{CKT,Kaz,Bro})
provides the homogeneous
transcendentality weight in all the orders of the expansion in $\ep$.
We see this property in our new piece of the result, i.e. in the $\ep^i$ terms with
$i=3,4,5$. Let us emphasize that this property is very useful when using PSLQ \cite{PSLQ}
because the number of constants that can be present in a result is essentially reduced.
In fact, we indeed arrived at the above result using PSLQ but could here proceed even without
the homogeneous transcendentality weight because, within the method of \cite{Lee:2009dh},
we obtained, for $I_{17}$, a very well convergent double series so that
we could obtain the accuracy of thousand of digits or more.
For the same reason, we can obtain higher terms of the expansion in $\ep$, i.e., $\ep^6$ etc.

If one pulls out a more sophisticated prefactor
\cite{CKT,Bro} (see the next equation)
the pure $\pi^2$ factors will not be present in the result.
We see that this property is satisfied indeed:
\bea
I_{17}=
(1-2\ep)^2\left(\frac{\Gamma (1-\ep)^2 \Gamma(1+\ep)}{\Gamma (2-2 \ep)}\right)^3
\biggl\{
20 \zeta (5)
&&\nn \\ &&  \hspace*{-78mm}
+\ep \left(68 \zeta (3)^2+\frac{10 \pi ^6}{189}\right)
\nn \\ &&  \hspace*{-78mm}
+\ep^2 \left(\frac{34 \pi ^4 \zeta (3)}{15}+450 \zeta(7)\right)
\nn \\ &&  \hspace*{-78mm}
 +\ep^3 \left(-\frac{9072}{5}\zeta(5,3)-2448 \zeta (3) \zeta (5)
+\frac{8519 \pi^8}{13500}\right)
\nn \\ &&  \hspace*{-78mm}
 +\ep^4 \left(-\frac{1292 \pi ^6 \zeta(3)}{189}-\frac{4640 \zeta (3)^3}{3}
 \right.
\nn \\ &&  \hspace*{-78mm}
\left.
+\frac{552 \pi ^4 \zeta (5)}{5}+\frac{88036\zeta (9)}{9}\right)
\nn \\ &&  \hspace*{-78mm}
+\ep^5 \left(42513 \zeta(8,2)-142178 \zeta (3) \zeta (7)
-73022 \zeta(5)^2
\right.
\nn \\ &&  \hspace*{-78mm}
\left.
-\frac{232 \pi ^4 \zeta (3)^2}{3}+\frac{593053 \pi^{10}}{187110}\right)
+\ldots\biggr\}\;.
\nn
\eea

Let us observe that in \cite{Brown:2008um}
it was proven that
the coefficients in the $\ep$-expansion of planar massless propagator diagrams
up to five loops
should be expressed in terms of multiple zeta values, while the non-planar graphs
may contain, in addition,
multiple sums with 6th roots of unity.
However, in the above result, only multiple zeta values are present.

\section{Conclusion}

Let us emphasize that in the present paper we used method of Ref.\cite{Lee:2009dh} in combination with several other methods. We already mentioned PSLQ. Then, as it was explained in \cite{Lee:2009dh,Lee:2010cg},
it is implied that the IBP reduction is solved for a given problem so that one can
obtain dimensional recurrence relations. To do this we used the code called {\tt FIRE} \cite{FIRE} and
the code based on \cite{Lee:2008tj} but, of course, one can use other methods.
Moreover, we use a sector decomposition \cite{BH,BognerWeinzierl,FIESTA} implemented in the code {\tt FIESTA}
\cite{FIESTA,FIESTA2} to determine the position and the order of the poles in the basic
stripe. To fix remaining constants in the homogenous solution of dimension recurrence relations
we apply the method of Mellin--Barnes representation \cite{MB1,MB2,books2}.

The method used in the present calculations looks quite promising and we expect it to be applied not
only in the problems discussed above but also in many other situations.

\vspace*{2mm}

\noindent
{\bf Acknowledgements.}
We would like to thank  D.J.~Broadhurst and K.G.~Chetyrkin for stimulating discussions.
This work was supported by the Russian Foundation for Basic Research through grant
08--02--01451.
V.S. appreciates the financial support of KIT through project
SFB/TR~9 for the participation in the workshop.

\end{document}